# The black holes are fictive objects

A. LOINGER

Dipartimento di Fisica dell'Università di Milano

Via Celoria, 16, 20133 Milano, Italy

ABSTRACT. – We prove that the anomalous form of Schwarzschild metric within the spatial domain bounded by Schwarzschild pseudosingular surface is a mathematical mishap, devoid of any physical meaning. As a special consequence, the portions of geodesics internal to the above domain are not physically interpretable. Accordingly, the theoretical notion of black hole describes only a fictive object.

## Introduction.

We show that there exists *a large class of static* (i.e. such that the $g_{mn}$'s are time independent) *and reversible* (in Eddington's terminology, i.e. such that $g_{01}=g_{02}=g_{03}=0$) *frames* for which the Schwarzschild pseudosingularity disappears. On the other hand, it is known that for the gravitational field of an electron there is no pseudosingularity, and the gravitational field of a fluid sphere is regular everywhere.

It follows that the anomalous form of Schwarzschild metric inside the spatial domain $\Im$ bounded by the pseudosingular surface is only a mathematical accident. This implies, in particular, the fundamental consequence that the portions of geodesics contained in $\Im$ do not admit of a physical interpretation. Thus the theoretical notion of black hole is destitute of physical reality.

We discuss in Appendix A the properties of an interesting *Bildraum*, and in Appendix B various conceptual items.

## 1.

Let us consider the four-dimensional line element $ds$ such that

$$(1) \quad ds^2 = \frac{r+(q-1)\alpha}{r+(q+1)\alpha} c^2 dt^2 - \frac{r+(q+1)\alpha}{r+(q-1)\alpha} dr^2 - [r+(q+1)\alpha]^2 d\omega^2,$$

where

$$(1') \quad d\omega^2 := d\vartheta^2 + \sin^2\vartheta \, d\varphi^2, \quad (0 \leq \vartheta \leq \pi; 0 \leq \varphi \leq 2\pi);$$

$q$ is a real parameter ($-\infty < q < +\infty$); $\alpha := GM/c^2$; $G$ is the constant of universal gravitation; $M$ is a mass concentrated in the origin of the spatial co-ordinate system ($r, \vartheta, \varphi$). For any value of $q$ we obtain a *new* space-time frame and a *new local* form of the solution of the static problem of a given mass point at rest; if $q$ is such that $1 < q < +\infty$, we have $0 \leq r < +\infty$, but for $q$ in the interval $-\infty < q \leq 1$ the line element (1) is conventionally interpretable only if $(1-q)\alpha < r < +\infty$.

If $q = -1$,

$$(2) \quad ds^2 = (1-2\alpha/r) c^2 dt^2 - (1-2\alpha/r)^{-1} dr^2 - r^2 d\omega^2,$$

i.e. the Einstein-Schwarzschild form (cf. e.g. Einstein, 1955, p.94). The label $r^*$ of the so-called critical radius $r_0$ is here equal to $2\alpha$. Various authors (cf. e.g. Landau-Lifchitz, 1966, p.360, Synge, 1971, p.270, Møller, 1972, p.439, Dirac, 1975, p.30, Kruskal, 1960 and Szekeres, 1960, Weyl, 1988, pp.250 and 346, and the literature quoted at p.349) have remarked that the singularity of (2) at the surface $r = 2\alpha$ is *not* an intrinsic one, and that it can be removed through the passage to proper non-static frames. Obviously, *an analogous conclusion holds also for* $r^* = (1-q')\alpha$, *when* $-\infty < q' \leq 1$.

If $q = 0$,

$$(3) \quad ds^2 = \frac{r-\alpha}{r+\alpha} c^2 dt^2 - \frac{r+\alpha}{r-\alpha} dr^2 - (r+\alpha)^2 d\omega^2,$$

i.e. the Fock's form, in harmonic co-ordinates (cf. e.g. Fock, 1964, p.209). The label $r^*$ of the critical radius $r_0$ is now equal to $\alpha$. (The use of the harmonic co-ordinates evidences clearly the close agreement of general relativity with Newton's theory).

If $q = 1/2$,

(4) $$ds^2 = \frac{r - \alpha/2}{r + 3\alpha/2} c^2 dt^2 - \frac{r + 3\alpha/2}{r - \alpha/2} dr^2 - (r + 3\alpha/2)^2 d\omega^2,$$

from which it follows that $r^* = \alpha/2$. (This value is also obtained in the so-called isotropic frame, cf. e.g. Pauli, 1958, p.248, Eddington, 1960, p.93, Landau-Lifchitz, 1966, Synge, 1971, Møller, 1972).

If $q = 1$,

(5) $$ds^2 = \frac{r}{r + 2\alpha} c^2 dt^2 - \frac{r + 2\alpha}{r} dr^2 - (r + 2\alpha)^2 d\omega^2,$$

and $r^*$ becomes equal to zero.

If $q = q'' > 1$, we have

(6) $$ds^2 = \frac{r + (q''-1)\alpha}{r + (q''+1)\alpha} c^2 dt^2 - \frac{r + (q''+1)\alpha}{r + (q''-1)\alpha} dr^2 - [r + (q''+1)\alpha]^2 d\omega^2,$$

and the pseudosingularity *disappears*, while the Newtonian singularity reveals itself – as in the other cases – in the prerelativistic limit: indeed, $g_{00} = [r+(q-1)\alpha]/[r+(q+1)\alpha] \approx 1 - 2\alpha/r$, for $-\infty < q < +\infty$. The above result is interesting since till now the disappearance of Schwarzschild pseudosingularity was *only* obtained by using appropriate non-static frames, see especially Kruskal, 1960 and Szekeres, 1960, who give a *global* treatment of Schwarzschild geometry.

## 2.

The train of thoughts of the inventors of the black holes can be summarized as follows. Relying on a well known Birkhoff's theorem, which asserts that all spherically symmetric solutions of the gravitational equations for empty space, going into the flat metric at infinity, are only different forms of Schwarzschild solution, they choose just this solution and emphasize the invariant meaning of the area $4\pi(2\alpha)^2$ of the surface $r=2\alpha$. Then, they utilize the *extended* non-static form of the solution (cf. e.g. Kruskal, 1960 and Szekeres, 1960) as a justification for investigating some hypothetical phenomena which could occur in $\Im := [0\leq r \leq 2\alpha]$ (black hole). (This region can be equivalently characterized by $0\leq r \leq(1-q')\alpha$, for $-\infty<q'\leq 1$, since, of course, the area of its surface is still equal to $4\pi(2\alpha)^2$).

### 3.

In general relativity *all* frames are conceptually on an equal footing, none of them can be theoretically privileged. Thus it is obviously correct to choose the Schwarzschild solution, provided that we bear in mind that the structure of Schwarzschild metric in the region $\Im$ is a mere mathematical mishap, with physically deceitful appearances. It is therefore illogical to take seriously the odd properties of the region $\Im$. On the other hand, from the physical point of view the non-static forms of the solution *cannot* be directly interpreted.

It is interesting to put in evidence the mathematical origin of the anomaly of the Schwarzschild form of the metric in the domain $\Im$. The resolution method by this author leads to the equation $rg_{00}= r+h$, where $h$ is an integration constant, *a priori* arbitrary. (The correspondence principle − asymptotic accord with Newton's theory − tells us that the correct physical value of $h$ is $-2\alpha$). If $h=0$, we have the Minkowskian metric, if $h>0$ we have a formula describing a fictive repulsive gravitational force. For $h\geq 0$ the interval $ds$ is well behaved, but already for an arbitrarily negative $h$ its behaviour becomes abnormal. **The right conclusion is that of the classic treatises on relativity: the Schwarzschild formula is *physically* interpretable only for $r>2\alpha$.**

Observe that if we start from a $g_{\vartheta\vartheta} = -(r+2\alpha+\varepsilon)^2$, with $\varepsilon>0$, we obtain the relation $(r+2\alpha+\varepsilon)g_{00} = r+k$; for $k=\varepsilon$, $g_{00} = (r+\varepsilon)/(r+2\alpha+\varepsilon) \approx 1-2\alpha/r$, in the Newtonian limit, and any difficulty has disappeared.

### 4.

Owing to the analytical complexity of the Einstein equations, the standard relativistic models of the gravitational collapse of a star are quite rudimentary.

"Einstein investigated the field of a system of many mass points, each of which is moving along a circular path, $r$=const., under the influence of the field created by the ensemble. If the axes of the circular paths are assumed to be oriented at random, the whole system or cluster is spherically symmetric. The purpose of the investigation was to find out whether the constituent particles can be concentrated toward the center so strongly that the total field exhibits a Schwarzschild singularity. The investigation showed that even before the critical concentration of particles is reached, some of the particles (those on the outside) begin to move with the velocity of light, that is along zero world lines. It is, therefore, impossible to concentrate the particles of the cluster to such a degree that the field has a [Schwarzschild] singularity". (From Bergmann, 1960, p.204). This Einsteinian model is more realistic than the standard model of the collapsing dust.

With an apparent paradox, several Newtonian treatments of the gravitational collapse are physically more adequate than the relativistic models (cf. e.g. Mc Vittie, 1964). But there is no room for the notion of black hole in the Newtonian mechanics (cf. e.g. Mc Vittie, 1978).

In the last thirty years innumerable papers on the black holes have been published. Sophistic notions have been coined as, e.g., "cosmic censorship", "naked (and clothed) singularities", "asymptotic predictability", etc., groundless hybridizations between general relativity and quantum theory have been invented, all the tools of topology and algebra have been used, with the result of producing a mass of metaphysical speculations, very far from Galilean physics.

ACKNOWLEDGEMENT

A useful discussion with my friend P. Bocchieri is gratefully acknowledged.

*———————————*

APPENDIX A

Let us re-write the interval (1) in the following way:

$$(A.1) \qquad ds^2 = \frac{r+(q-1)\alpha}{r+(q+1)\alpha} c^2 dt^2 - \frac{r+(q+1)\alpha}{r+(q-1)\alpha} d\sigma^2,$$

where

$$(A.2) \qquad d\sigma^2 := dr^2 + [(r+q\alpha)^2 - \alpha^2] d\omega^2;$$

$d\sigma$ is the line element in a three-dimensional auxiliary space $\mathcal{L}(q)$ (a *Bildraum*), which is almost Euclidean; for $q=0$ it coincides with Fock's auxiliary "conformal space". In $\mathcal{L}(q')$ the labels $r^*(q')$, $-\infty < q' \leq 1$, of the critical radius $r_0$ acquire an elementary *metrical* meaning; indeed, $\int_0^{r^*(q')} dr = r^*(q')$ is the length of the line segment $[r=0, r= r^*(q')]$. Therefore in the *Bildraum* $\mathcal{L}(q')$ the black hole of mass $M$ has a Euclidean radius which depends on the parameter $q'$, while the area of its surface is equal to zero. With implicit reference to the auxiliary space $\mathcal{L}(q'=0)$, Fock gives in his book (1964) the numerical values of the critical radii of several celestial bodies.

APPENDIX B

*i*) **An objection**: From the intrinsic and global standpoint of differential geometry the geodesics are interrupted only at the (true) singularities, in our case only at $r=0$ (Newtonian singularity). In other terms, any space-time frame without the region $\Im$ of invariant surface $4\pi(2\alpha)^2$ would be geodesically incomplete. ***The answer***: This is true from the purely mathematical point of view, nevertheless the crucial point is the following: a straightforward *physical* interpretation is not possible in non-static frames. Further, we remark that the static frames of sect.**1** are also *reversible*, i.e. such that $g_{14}=g_{24}=g_{34}=0$; in frames of this kind "the time will be reversible […]; this renders the application of the name "time" to $x_4$ more just […]". (A.S. Eddington, 1960, p.81). Now, in the static and reversible Schwarzschild metric the portions of geodesics within the domain $\Im$ have only a mathematical sense. Summing up, all frames are on an equal footing from the abstract point of view, but some of them are "more equal" than others so far as the physical interpretation is concerned. This fundamental consideration was efficaciously emphasized, e.g., by Eddington and Fock.

*ii*) Eq. (1) of sect.**1** is simply obtained from Schwarzschild metric by shifting the co-ordinate $r$ by the constant $(q+1)\alpha$. In this trivial way we see, in particular, that there are also static and reversible frames for which the Schwarzschild pseudosingularity disappears. (This conclusion is perfectly in the spirit of the introductory remarks of the paper by Szekeres, 1960). Only if the new $r$ − which is obviously non-negative − is such that $(1−q)\alpha < r < +\infty$, when $−\infty < q \leq 1$, the line element d$s$ is physically interpretable, according to the criterion adopted, e.g., by Einstein, Weyl, Pauli, Eddington and Fock. Thus, with reference to harmonic co-ordinates, $(q=0)$, $r \geq 0$, $\vartheta$, $\varphi$, Fock writes (1964, p.210 and p.214): "By its physical nature $\rho \ [:= \sqrt{r^2 - \alpha^2}$ ; cf. eq. (57.44)] must be positive and therefore the range of variation of $r$ is $r > \alpha$".

*iii*) As it is well known, the gravitational field of an electron is given by (see e.g. A.S. Eddington, 1960, p.185):

(B.1) $$ds^2 = -\gamma^{-1}dr^2 - r^2 d\omega^2 + \gamma dt^2 \ ,$$

with

(B.2) $$\gamma := 1 - \frac{2m}{r} + \frac{4\pi\varepsilon^2}{r^2} \, ,$$

where $m$ (which coincides with our $\alpha$) and $4\pi\varepsilon$ (the electric charge) are constants of integration. Put $a:=2\pi\varepsilon^2/m$. Eddington writes: "When $r$ is diminished the value of $\gamma$ […] decreases to a minimum for $r=2a$, and then increases continually becoming infinite at $r=0$. There is no singularity in the electromagnetic and gravitational fields except at $r=0$". An analogous result holds for any charged mass point for which $m<2a$. This example shows clearly that the Schwarzschild pseudosingularity is a mere mathematical accident.

*iv*) It is also significant that for the model of a fluid sphere the Einstein field is regular both externally and *internally* to the sphere.

*v*) Einstein's negative judgements on the physical meaning of the singularities of any kind are well known. For instance, Einstein wrote (1955, p.129): "One may not […] assume the validity of the equations for very high density of field and matter". And in a paper with N. Rosen (1935) just dedicated to the Schwarzschild singularity: " Every field theory, in our opinion, must […] adhere to the fundamental principle that singularities of the field are to be excluded".

*vi*) So far we know only the solutions of the Einstein equations giving the fields produced by a single mass distribution with radial symmetry about a centre, or with rotational symmetry about an axis (Kerr's solutions). "There is, therefore, no way of asserting […] that a black hole could be a component of a close binary system or that two black holes could collide. An existence theorem would first be needed to show that Einstein's field equations contained solutions which described such configurations". (Mc Vittie, 1978).

## PARERGON

Abrams (1979) has demonstrated that the relativistic standard EXTERIOR metric of a mass point at rest is diffeomorphic to the ORIGINAL (non-standard !) Schwarzschild's solution of the same problem (1916), *which is perfectly well-behaved in the whole space-time*. Clearly this significant result strengthens the thesis according to which the notion of black hole is only the insane product of a wishful fiction. On the other hand, the Great Spirits who created and developed the Relativity (Einstein, Weyl, Levi-Civita, Eddington, Pauli, Fock, ...) were of the unanimous conviction that only the EXTERIOR part of the standard solution of Einstein's equations for a point mass has a real physical meaning.

An interesting discussion with Dr. S. Antoci, who sent me a copy of the above paper by Abrams, is gratefully acknowledged.

*————————————*